# Fully on-chip photonic turnkey quantum source for entangled qubit/qudit state generation

Hatam Mahmudlu,[1,2,3] Robert Johanning,[1,2,3] Anahita Khodadad Kashi,[1,2,3] Albert van Rees,[4] Jörn P. Epping,[5,6] Raktim Haldar,[1,2,3,*] Klaus-J. Boller,[4] and Michael Kues[1,2,3,†]

[1]*Institute of Photonics, Leibniz University Hannover,*
*Nienburger Straße 17, 30167 Hannover, Germany*
[2]*Hannover Centre for Optical Technologies, Leibniz University Hannover,*
*Nienburger Straße 17, 30167 Hannover, Germany*
[3]*Cluster of Excellence PhoenixD (Photonic, Optics,*
*and Engineering – Innovation Across Disciplines),*
*Leibniz University Hannover, Hannover, Germany*
[4]*Laser Physics and Nonlinear Optics, Mesa+Institute for Nanotechnology,*
*Department for Science and Technology, University of Twente,*
*PO Box 217, Enschede, 7500 AL, The Netherlands*
[5]*QuiX Quantum, 7521, AN Enschede, The Netherlands*
[6]*Lionix Intl., 7521, AN Enschede, The Netherlands*

Integrated photonics has recently become a leading platform for the realization and processing of optical entangled quantum states in compact, robust and scalable chip formats with applications in long-distance quantum-secured communication, quantum-accelerated information processing, and non-classical metrology. However, the quantum light sources developed so far have relied on external bulky excitation lasers making them impractical, not reproducible prototype devices, hindering scalability and the transfer out of the lab into real-world applications. Here we demonstrate a fully integrated quantum light source, which overcomes these challenges through the combined integration of a laser cavity, a highly efficient tunable noise suppression filter ($>55\,$dB) exploiting the Vernier effect, and a nonlinear microring for entangled photon pair generation through spontaneous four-wave mixing. The hybrid quantum source employs an electrically-pumped InP gain section and a $Si_3N_4$ low-loss microring filter system, and demonstrates high performance parameters, i.e., a pair emission over four resonant modes in the telecom band (bandwidth $\sim1\,$THz), and a remarkable pair detection rate of $\sim620\,$Hz at a high coincidence-to-accidental ratio of $\sim80$. The source directly creates high-dimensional frequency-bin entangled quantum states (qubits/qudits), verified by quantum interference measurements with visibilities up to $96\,\%$ (violating Bell-inequality) and by density matrix reconstruction through state tomography showing fidelities of up to $99\,\%$. Our approach, leveraging a hybrid photonic platform, enables commercial-viable, low-cost, compact, light-weight, and field-deployable entangled quantum sources, quintessential for practical, out-of-lab applications, e.g., in quantum processors and quantum satellite communications systems.

Integrated quantum photonics, due to the availability of mature fabrication technologies, sophisticated information processing tools at room temperature, and the robustness to decoherence, is considered one of the most compelling approaches for the development of future nonclassical technologies[1–3]. The contemporary semiconductor industry allows to fabricate highly compact and robust, mass-producible on-chip photonic devices[4], where major advances have demonstrated the capability of generating and controlling entangled quantum optical states on-chip[5,6], which is of paramount importance for quantum key distribution (QKD)[7,8] and quantum information processing (QIP)[8–17,19–21]. Particularly, existing on-chip realizations exhibit two-photon interference[9,10], teleportation[11], quantum walks[12], boson sampling[13], QIP involving multiple entangled photons[14–16], error-protected qubits[17], noise-resilient qudits[8,19,20], cluster states[21], and eventually the development of quantum photonic processors[16,20] are just a few examples considered to be the watershed of integrated quantum photonic technologies.

Notably, all previous on-chip entangled quantum photonic sources[8,22] and the pioneering on-chip demonstrations of nonclassical functionalities were until now dependent on external excitation lasers, thereby making these systems overall non-reproducible, bulky, impractical, and thus unsuitable for out-of-lab use as well as production at large scale[8,12–17,19–21]. To address this drawback, it is critical to realize a fully integrated quantum light source of entangled photons, which will allow all stages of QIP to be on a single chip, bringing ultimately the required stability and scalability[3,4]. However, to date, the major technical challenge inhibiting a turnkey quantum light system was to integrate a stable and tunable laser together with a filter, which is crucial to eliminate noise that suppresses quantum phenomena[8], and a nonlinear parametric source of entangled photons (i.e., creating signal and idler photons through spontaneous parametric effects)[4,23,24].

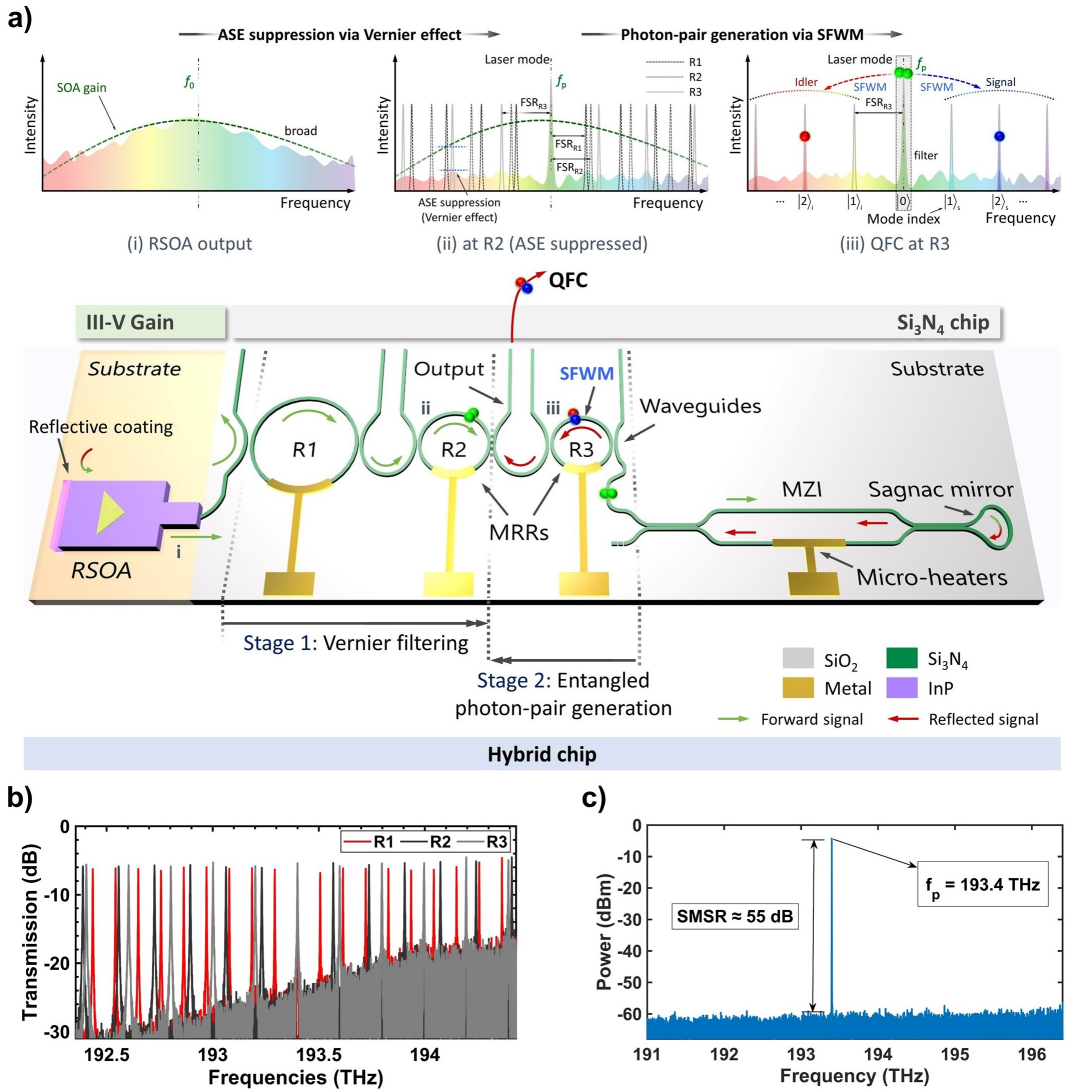

FIG. 1. **a) Laser-integrated photonic quantum light source of frequency-bin entangled photon pairs:** An electrically pumped InP RSOA providing optical gain is coupled to a $Si_3N_4$ chip containing a three microring (R1, R2, and R3) Vernier filter and a Mach-Zehnder interferometer (MZI) with a loop mirror. The free spectral range (FSRs) of R1, R2, and R3 are $\sim 107$ GHz, $\sim 167$ GHz, and $\sim 199$ GHz, respectively. The spectral positions of the ring resonances can be adjusted using electrically driven integrated microheaters to provide a single frequency filter in the gain bandwidth through the Vernier effect. Together with the gain section and the cavity built by the end mirror and loop mirror, the system allows laser operation on this filter line. The backward propagating laser field acts as the excitation signal for a spontaneous four-wave mixing process within the third ring R3 producing a quantum frequency comb. **b)** Measured transmission spectra of $Si_3N_4$ microrings R1, R2, and R3. **c)** Classical output spectrum showing the laser line at a frequency of 193.4 THz with a high SMSR of $> 55$ dB.

The absence of a unique material platform that is superior in all quantum photonic functionalities (i.e., low-loss guiding, filtering, efficient parametric generation of entangled photon pairs, and their active manipulation) while providing at the same time lasing gain, mainly impedes monolithic integration. A broad range of photonic platforms, e.g., silica ($SiO_2$)[14], silicon-on-insulator (SOI)[9], silicon nitride ($Si_3N_4$)[8,10], lithium niobate[6], gallium arsenide (GaAs), indium arsenide (InAs), indium phosphide (InP)[22] with a potpourri of exotic compounds[25] and 2D materials[26] have been investigated in the context of quantum processing and parametric photon pair generation. Most commonly, low-loss indirect or high-bandgap materials with high refractive index (e.g., Si, $SiO_2$, $Si_3N_4$) are preferred for light guiding, signal processing as well as for efficient spontaneous parametric effects and thus entangled photon pair genera-tion, whereas direct-bandgap III-V semiconductors (e.g., InAs, GaAs, InP) are suitable for optical gain and lasing. Unfortunately, the fabrication technique/process flow for each group of materials is different and often not compatible with each other[24]. Thus, the hybrid integration of the excitation laser with a parametric photon source into a combined photonic circuit, drawing the utmost advantages of different materials while avoiding the deficiencies, is considered the key step toward a fully-integrated on-chip quantum device[23,24].

While in addition to parametric spontaneous sources, also semiconductor quantum-dots (QDs) can emit single[24] and entangled photons[27] in compact forms, they are yet driven by external fields. Electrically-pumped QD-lasers containing QDs have the potential to act as the primary excitations to generate entangled photons on-chip. Nevertheless, it is demanding to integrate QDs with a fully operational quantum photonic circuit due to random growths of QDs on wafers, the presence of charge-noise, the lack of precise emission-wavelength and the need of cryogenic cooling to ensure indistinguishable, narrow band ($\sim$MHz), and deterministic single-photon output. Therefore, the QDs must be connected with the rest of the circuit either by optical fibers, which makes the device bulky, or by wafer bonding, where loss of the single photon signal due to misalignment and refractive index mismatch at the circuit interface becomes unavoidable[24].

The emission from InP quantum wells covers the entire C-band making InP a perfect candidate as a semiconductor optical amplifier (gain medium) to realize lasers that would drive on-chip quantum light sources. However, the intrinsic waveguide losses in such amplifiers and also in InP passive waveguides are high. This imposes a short photon lifetime of the laser cavity causing a high spectral linewidth. A solution to this is extending the photon lifetime with hybrid integration of waveguide feedback circuits based on $Si_3N_4$ waveguides[4], which offers a broad transparency window covering the entire C-band with remarkably low loss. Moreover, $Si_3N_4$ has a high nonlinear refractive index ($n_2 \approx 2.4 \times 10^{-19}$ m$^2$/W), moderately high mode confinement, low material dispersion, near to zero Raman scattering[29,30] and owing to the large bandgap, absence of two-photon absorption (TPA) in the telecom wavelengths range. Absence of TPA in $Si_3N_4$ permits to operate in a high-power regime, effectively reducing the intrinsic laser linewidth to the sub-kHz range. These characteristics make $Si_3N_4$ particularly suitable for nonlinear and quantum photonic applications[8,29–31]. Over the past few years, compact, high-Q $Si_3N_4$ microresonators[6,33] with ultra-narrow resonant linewidth and extremely small mode volumes were employed in the classical domain for the generation of low-threshold Kerr frequency combs[8,29–31], as well as in the quantum domain for the realization of squeezed state of light[34] and entangled photon pairs via spontaneous four-wave mixing[8]. Very recently mode-locked micro combs have been reported using laser-integrated photonic chips[5,36,37]. Here, for the first time, we address the afore-mentioned challenges related to the full integration and demonstrate a hybrid, integrated turnkey quantum light source emitting frequency-bin entangled qubit and qudit states. The system is illustrated in Fig. 1a and consists of an electrically pumped III-V reflective semiconductor optical amplifier (RSOA) based on InP providing optical gain, which is coupled to a low-loss $Si_3N_4$ waveguide feedback circuit. The reflective coating at the amplifier side and a Sagnac mirror at the opposite end together form a laser cavity. After the gain-stage, three microring resonators (MRRs) with different radii (R1>R2>R3) are placed in a cascaded fashion, creating a Vernier filter (see Fig 1b), which is tunable through the incorporated microring phase shifters. The combination of low propagation loss and tight guiding through high index contrast in $Si_3N_4$ waveguides enables the realization of MRRs with small radii, vital components to realize Vernier filters with large FSRs. This is followed by a Mach-Zehnder interferometer and the Sagnac loop mirror. Several measures have been taken, e.g., waveguide tapering and tilting at the $Si_3N_4$ input, anti-reflection coatings at the interface, etc. to mitigate the back-reflection spawning from the refractive index mismatch between the InP and $Si_3N_4$ circuit interface[4] (see Methods). Finally, our material choice (InP-$Si_3N_4$) in combination with Vernier filtering led to a single frequency laser emission from the gain medium (see Fig. 1c), ideal for exciting only a particular longitudinal mode of the microring resonator. The intra-cavity laser field (set at 193.4 THz (1550 nm)) back-reflected from the Sagnac mirror acts as an excitation field for a nonlinear spontaneous four-wave mixing (SFWM) process within the third MRR (R3). This in turn produced correlated signal/idler photon pairs in the frequencies symmetric to the excitation field (i.e., $f_p = 193.4$ THz) forming a bi-photon quantum frequency comb (QFC). The QFC excited by a continuous-wave source generally dwells in high-dimensional Hilbert-space since the idler and the signal photons are in superposition of multiple frequency modes owing to the energy conservation of SFWM[8]. The advantage of this configuration is twofold: first, the parametric quantum source is directly incorporated in the laser cavity such that the excitation field is locked to the Vernier resonance for stable operation, second, the Vernier filter is directly suppressing the amplified spontaneous emission (ASE) from the InP gain on the relevant frequencies.

In order to analyze the quantum properties of the source, we performed photon correlation measurements on several resonance pairs (see Fig. 2a) at the generation stage 2 (see Fig. 1 a-b) via the setup configuration and measurement procedure described in the Methods. Our measurements reveal coincidence counts (Fig. 2b) demonstrating the generation of signal/idler photon pairs (see Fig. 2c and d) and the creation of a bi-photon QFC. From the full width at half maximum of the measured coincidence histogram, shown for one resonance pair in Fig. 2b, a photon coherence time of approximately 150 ps was deduced, being in good agreement with the measured resonance linewidth of 5.1 GHz. Note that, for the photon pair generation, we explicitly chose the third MRR (R3) excited by the backward propagating field instead of the forward propagation, due to a major advantage: The noise, largely emerging from the ASE present in the system, is intrinsically suppressed due to the chip configuration. Thus, as it is seen in Fig. 1a, the noise emanating from the gain medium experiences a drop at the Vernier filtering stage, in the forward propagation (Stage 1 in Fig. 1a) also manifested by the MRR transmission in Fig. 1b. The measurements revealed four resonance pairs with major coincidence counts as shown in Fig. 2c, revealing a ~1 THz photon pair generation bandwidth. Such a moderate photon-pair generation bandwidth is the consequence of the normal dispersion regime, i.e., a positive group delay dispersion (GDD) at the excitation wavelength. In addition, we measured a CAR of approximately 80 per signal/idler resonance pair, (see Fig. 2c). A system with such performance can be readily implemented in many quantum scenarios[8]. Furthermore, for a detected excitation field power of around 3.9 mW, we determined a detection rate of ~620 Hz, as well as a calculated coincidence rate of ~8200 Hz at the device output, i.e., considering the insertion losses of the filter as well as the quantum efficiencies of the detectors.

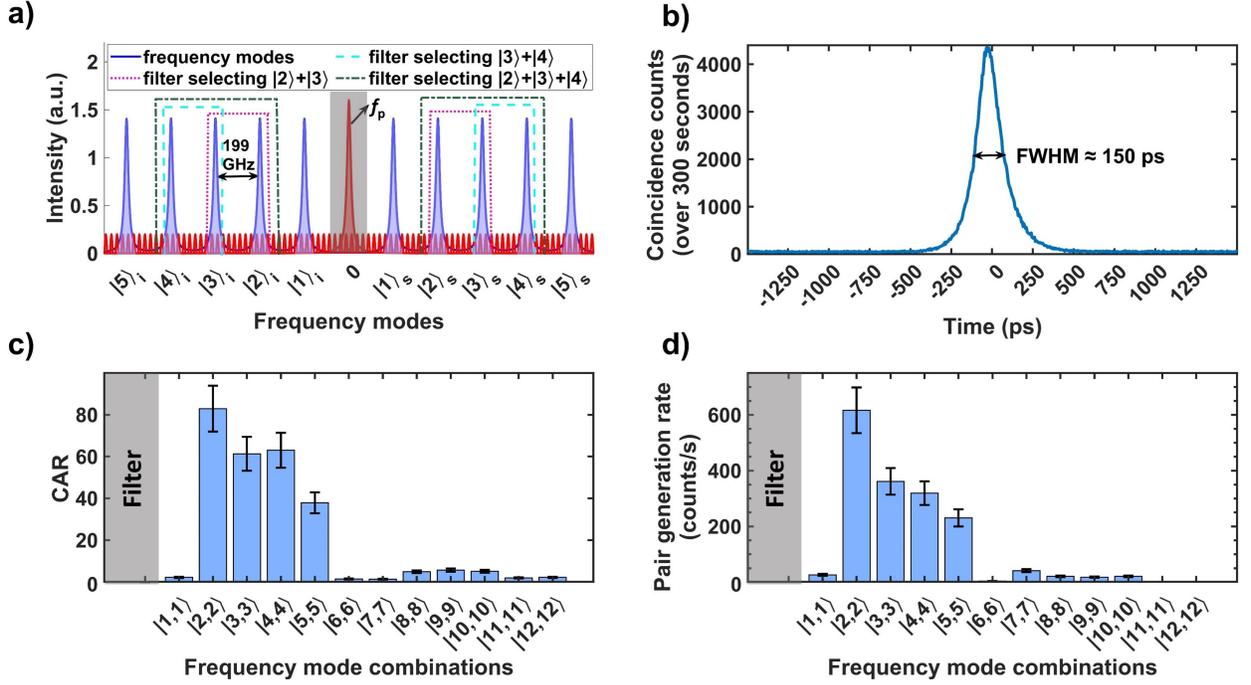

FIG. 2. **Performance characteristics of the quantum light source. a)** Schematic diagram for the description of the QST and Bell test measurements. Photon pairs created through the SFWM process reside on spectrally symmetric idler and signal resonances. $f_p$ denotes the pump frequency. **b)** Histogram of time difference between coincidence counts collected for 300 seconds showing photon pair correlations in the emission time. The full width at half maximum of the peak is 150 ps matching with the 5.1 GHz resonance linewidth. **c)** Coincidence-to-accidental ratio (CAR) for respective frequency pairs for a detected excitation power of 3.9 mW. **d)** Detection rate, i.e., the number of the coincidence counts per second, for the respective frequency pairs. The spectral response of the 200 GHz wide notch filter (NF), for rejection of the excitation field used is illustrated schematically with transparent grey color in all plots.

The photon source directly emits frequency-bin entangled high dimensional states, which was verified by Bell test and quantum state tomography measurements. For this, a frequency shift via an electro-optic modulator (EOM) was used to superimpose individual frequency modes to implement quantum interference and projection measurements (see Methods). We defined the two-qubit (two dimensional) and two-qutrit (three dimensional) states by selecting the second, third, and the fourth frequency mode for both the signal and idler photon (see Fig. 2a). As these frequency pairs have similar state amplitudes (see Fig. 2c and Fig. 2d), we expect that maximally entangled Bell-states are generated[8]. The quantum interference patterns for the qubit and qutrit states are shown in Fig. 3 (i), yielding raw visibilities of $V_{2-3}^{(2)} = 0.975 \pm 0.002$ and $V_{3-4}^{(2)} = 0.976 \pm 0.002$ for qubits and $V_{2-3-4}^{(3)} = 0.968 \pm 0.004$ for qutrits (without background subtraction). Here the superscripts of the symbols within the first brackets represent the dimension and the subscripts denote the frequency mode numbers. These values violate the respective Bell inequalities for qubits and qutrits by greatly exceeding the threshold values of ~0.7071 and ~0.7746, respectively (see Methods)[8]. Furthermore, we reconstructed the density matrices corresponding to the generated quantum states by QST (see Methods). The absolute and argument values of the respective density matrices can be seen in Fig. 3 (ii). It turns out that the source generates Bell states, which were measured with excellent fidelities of $F_{2-3}^{(2)} = 0.988 \pm 0.01$ and $F_{3-4}^{(2)} = 0.981 \pm 0.01$ for qubits and $F_{2-3-4}^{(3)} = 0.806 \pm 0.02$ for qutrits. The latter shows an improved fidelity of $F_{2-3-4}^{(3),\ \mathrm{noise}} = 0.857 \pm 0.02$ by taking 6% linear noise into account, while the qubit case does not show a significant improvement by this approach (see Methods). Further, the 2D density matrices show a violation of the CHSH-inequality with up to $S = 2.793$ which is in good agreement with the values derived from the Bell test with up to $S = 2.760$ (see Methods). These measurements prove the excellent quality of the qubit/qudit pairs and in turn their large potential for applications in quantum technologies.

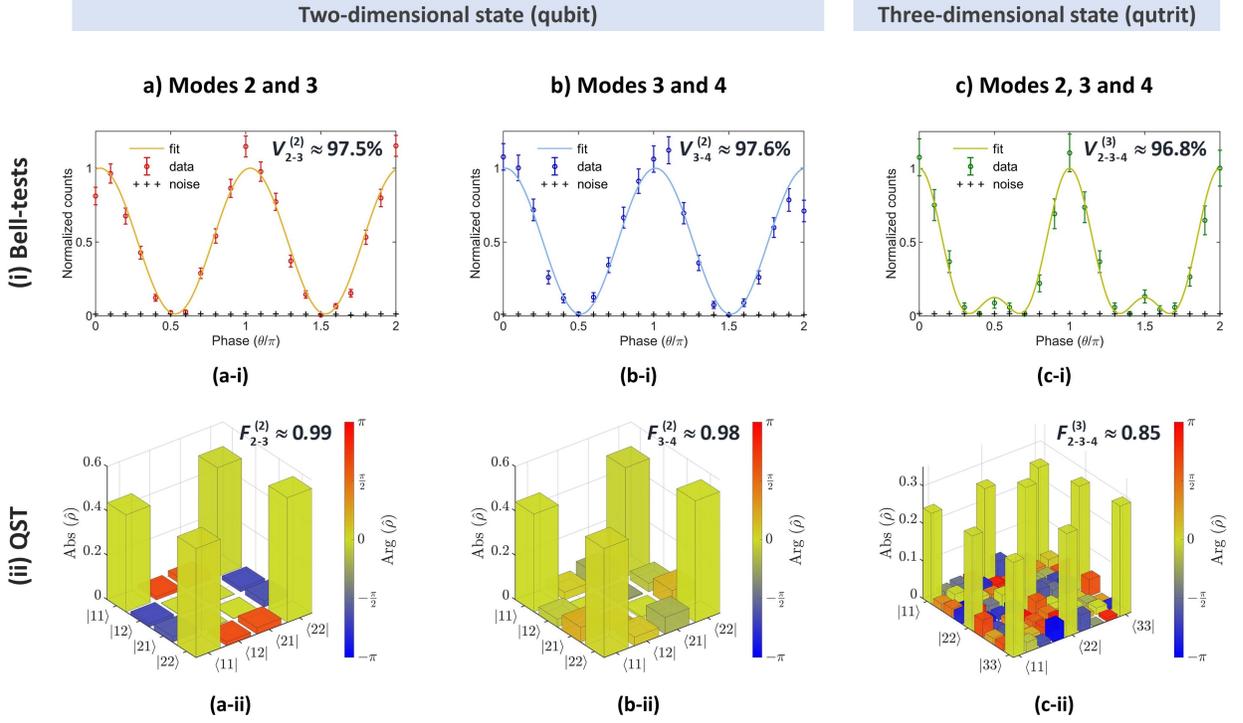

FIG. 3. **Bell inequality violation and quantum state tomography of frequency-entangled qubit and qutrit states.** **First line (i),** Bell test characterization of qubit and qutrit states. The error bars show the root of the absolute count numbers; the error of the noise background is too small to be plotted. **Second line (ii),** quantum state tomography for qubit and qutrit QFC states. **First column a),** two qubit state (composed of 2nd and 3rd resonances). **Second column b),** two qubit state (composed of 3rd and 4th resonances). **Third column c),** two qutrit state (composed of 2nd, 3rd and 4th resonances). The fidelity was calculated with respect to the three-dimensional Bell state mixed with 6 % linear noise.

Our measurements on qutrits[10] do not yet exhaust the maximum number of ring resonances. The maximum number of signal/idler pairs with efficient pair generation rates is limited by the SFWM bandwidth which depends on the dispersion of the $Si_3N_4$ waveguide. Bell-test and the QST measurements are limited by the maximum number of frequency modes that can be mixed, which is determined by the ratio of the FSR and the EOM driving frequency (see Methods). In this sense, the designed

FSR of the ring resonator during the photon pair generation could be decreased which enables to increase the number of frequency modes for mixing, and in turn our access to the effective dimensions. However, too much decrease in the resonator FSR could decrease the total Vernier FSR to the extent leading to the violation of single pump frequency oscillation in the cavity by creating more pump modes within the spectral bandwidth used in our experiments (C-band). Therefore, according to the Vernier principle, this procedure is also limited and needs careful handling to prevent extra possible challenges such as the need for filtering out more pump modes.

In conclusion, building on the latest advances in hybrid integration technologies, we presented for the first time an electrically-pumped laser-integrated versatile photonic quantum light source entirely on-chip. The sub-millimeter-sized chip accommodates a laser cavity and a Vernier filtering stage capable of generating high-quality, low-noise entangled photon pairs residing in a high-dimensional Hilbert space. We attained a remarkably high pair generation rate of 8200 counts/s, maximum CAR of $> 80$ and CAR of at least ($\sim 40$) for all the four resonance pairs, spanning $\sim 1$ THz around 1550 nm. The photon generation rate and the generation bandwidth can be increased by suitably tailoring and flattening the GDD of the $Si_3N_4$ waveguide. Further reduc-tion in surface roughness, waveguide-loss, enhancement in MRRs' Q-factors[33] and in turn ASE suppression can lead to a better CAR. The state fidelity close to unity (0.99) and the quantum interference visibility (96 %) are superior to previous demonstrations with external lasers[8] yet here within a much smaller footprint.

The proposed design empowers intrinsic long-term subwavelength stability towards high complexity with a high number of sources, and seamless integration with subsequent, upscaled quantum processing circuits. Therefore, our approach can be exploited in merely every presented on-chip QIP scheme, e.g., boson sampling, teleportation, certified quantum random number generation. Future steps would focus on including this source in fully-integrated programmable quantum processors where piezo or electro-optic phase shifters can be used to realize the desired state generation, e.g., hyper-entangled or a cluster-states[21] and state process-ing. In this context, in addition to the frequency-bin encoding, a multiplexing of the presented source, easily achievable, can realize widely-used path-encoded quantum states. We thus envisage our end-to-end integrated, tiny quantum light source to be an elementary constituent of densely-packed programmable photonic quantum processors[2,16,20] in near future. Moreover, this portable high-performing multi-level ($M$-ary) entangled photon source is of high importance for quantum communications. It can readily serve as a practical, mass-producible building-block to drive quantum networks, with the possibility to use frequency-encoding, perfectly suited for state-of-the-art telecommunications components[3]. By significantly reducing the payload, while being robust and stable, it can be perfect for applications in quantum satellite communications. Finally, this work could be considered as one step forward toward achieving quantum supremacy on-chip using photons[39].

---

* raktim.haldar@iop.uni-hannover.de

† michael.kues@iop.uni-hannover.de

1 O'brien, J., Furusawa, A. & Vučković, J. Photonic quantum technologies. *Nature Photonics.* **3**, 687-695 (2009)

2 Carolan, J., Harrold, C., Sparrow, C., Martín-López, E., Russell, N., Silverstone, J., Shadbolt, P., Matsuda, N., Oguma, M., Itoh, M. & Others Universal linear optics. *Science.* **349**, 711-716 (2015)

3 Wang, J., Sciarrino, F., Laing, A. & Thompson, M. Integrated photonic quantum technologies. *Nature Photonics.* **14**, 273-284 (2020)

4 Pelucchi, E., Fagas, G., Aharonovich, I., Englund, D., Figueroa, E., Gong, Q., Hannes, H., Liu, J., Lu, C., Matsuda, N. & Others The potential and global outlook of integrated photonics for quantum technologies. *Nature Reviews Physics.* pp. 1-15 (2021)

5 Shadbolt, P., Verde, M., Peruzzo, A., Politi, A., Laing, A., Lobino, M., Matthews, J., Thompson, M. & O'Brien, J. Generating, manipulating and measuring entanglement and mixture with a reconfigurable photonic circuit. *Nature Photonics.* **6**, 45-49 (2012)

6 Jin, H., Liu, F., Xu, P., Xia, J., Zhong, M., Yuan, Y., Zhou, J., Gong, Y., Wang, W. & Zhu, S. On-chip generation and manipulation of entangled photons based on reconfigurable lithium-niobate waveguide circuits. *Physical Review Letters.* **113**, 103601 (2014)

7 Wei, K., Li, W., Tan, H., Li, Y., Min, H., Zhang, W., Li, H., You, L., Wang, Z., Jiang, X. & Others High-speed measurement-device-independent quantum key distribution with integrated silicon photonics. *Physical Review X.* **10**, 031030 (2020)

8 Lu, X., Li, Q., Westly, D., Moille, G., Singh, A., Anant, V. & Srinivasan, K. Chip-integrated visible–telecom entangled photon pair source for quantum communication. *Nature Physics.* **15**, 373-381 (2019)

9 Silverstone, J., Bonneau, D., Ohira, K., Suzuki, N., Yoshida, H., Iizuka, N., Ezaki, M., Natarajan, C., Tanner, M., Hadfield, R. & Others On-chip quantum interference between silicon photon-pair sources. *Nature Photonics.* **8**, 104-108 (2014)

10 Zhang, X., Bell, B., Mahendra, A., Xiong, C., Leong, P. & Eggleton, B. Integrated silicon nitride time-bin entanglement circuits. *Optics Letters.* **43**, 3469-3472 (2018)

11 Metcalf, B., Spring, J., Humphreys, P., Thomas-Peter, N., Barbieri, M., Kolthammer, W., Jin, X., Langford, N., Kundys, D., Gates, J. & Others Quantum teleportation on a photonic chip. *Nature Photonics.* **8**, 770-774 (2014)

12 Peruzzo, A., Lobino, M., Matthews, J., Matsuda, N., Politi, A., Poulios, K., Zhou, X., Lahini, Y., Ismail, N., Wörhoff, K. & Others Quantum walks of correlated photons. *Science.* **329**, 1500-1503 (2010)

13 Paesani, S., Ding, Y., Santagati, R., Chakhmakhchyan, L., Vigliar, C., Rottwitt, K., Oxenløwe, L., Wang, J., Thompson, M. & Laing, A. Generation and sampling of quantum states of light in a silicon chip. *Nature Physics.* **15**, 925-929 (2019)

14 Matthews, J., Politi, A., Stefanov, A. & O'brien, J. Manipulation of multiphoton entanglement in waveguide quantum circuits. *Nature Photonics.* **3**, 346-350 (2009)

15 Reimer, C., Kues, M., Roztocki, P., Wetzel, B., Grazioso, F., Little, B., Chu, S., Johnston, T., Bromberg, Y., Caspani, L. & Others Generation of multiphoton entangled quantum states by means of integrated frequency combs. *Science.* **351**, 1176-1180 (2016)

16 Arrazola, J., Bergholm, V., Brádler, K., Bromley, T., Collins, M., Dhand, I., Fumagalli, A., Gerrits, T., Goussev, A., Helt, L. & Others Quantum circuits with many photons on a programmable nanophotonic chip. *Nature.* **591**, 54-60 (2021)

17 Vigliar, C., Paesani, S., Ding, Y., Adcock, J., Wang, J., Morley-Short, S., Bacco, D., Oxenløwe, L., Thompson, M., Rarity, J. & Others Error-protected qubits in a silicon photonic chip. *Nature Physics.* **17**, 1137-1143 (2021)

18 Kues, M., Reimer, C., Roztocki, P., Cortés, L., Sciara, S., Wetzel, B., Zhang, Y., Cino, A., Chu, S., Little, B. & Others On-chip generation of high-dimensional entangled quantum states and their coherent control. *Nature.* **546**, 622-626 (2017)

19 Wang, J., Paesani, S., Ding, Y., Santagati, R., Skrzypczyk, P., Salavrakos, A., Tura, J., Augusiak, R., Mančinska, L.,

Bacco, D. & Others Multidimensional quantum entanglement with large-scale integrated optics. *Science.* **360**, 285-291 (2018)

[20] Chi, Y., Huang, J., Zhang, Z., Mao, J., Zhou, Z., Chen, X., Zhai, C., Bao, J., Dai, T., Yuan, H. & Others A programmable qudit-based quantum processor. *Nature Communications.* **13**, 1-10 (2022)

[21] Adcock, J., Vigliar, C., Santagati, R., Silverstone, J. & Thompson, M. Programmable four-photon graph states on a silicon chip. *Nature Communications.* **10**, 1-6 (2019)

[22] Horn, R., Abolghasem, P., Bijlani, B., Kang, D., Helmy, A. & Weihs, G. Monolithic source of photon pairs. *Physical Review Letters.* **108**, 153606 (2012)

[23] Xiang, C., Guo, J., Jin, W., Wu, L., Peters, J., Xie, W., Chang, L., Shen, B., Wang, H., Yang, Q. & Others High-performance lasers for fully integrated silicon nitride photonics. *Nature Communications.* **12**, 1-8 (2021)

[24] Elshaari, A., Pernice, W., Srinivasan, K., Benson, O. & Zwiller, V. Hybrid integrated quantum photonic circuits. *Nature Photonics.* **14**, 285-298 (2020)

[25] Moody, G., Sorger, V., Blumenthal, D., Juodawlkis, P., Loh, W., Sorace-Agaskar, C., Jones, A., Balram, K., Matthews, J., Laing, A. & Others Roadmap on integrated quantum photonics. *ArXiv Preprint ArXiv:2102.03323.* (2021)

[26] Shiue, R., Efetov, D., Grosso, G., Peng, C., Fong, K. & Englund, D. Active 2D materials for on-chip nanophotonics and quantum optics. *Nanophotonics.* **6**, 1329-1342 (2017)

[27] Versteegh, M., Reimer, M., Jöns, K., Dalacu, D., Poole, P., Gulinatti, A., Giudice, A. & Zwiller, V. Observation of strongly entangled photon pairs from a nanowire quantum dot. *Nature Communications.* **5**, 1-6 (2014)

[28] Fan, Y., Rees, A., Slot, P., Mak, J., Oldenbeuving, R., Hoekman, M., Geskus, D., Roeloffzen, C. & Boller, K. Hybrid integrated InP-Si 3 N 4 diode laser with a 40-Hz intrinsic linewidth. *Optics Express.* **28**, 21713-21728 (2020)

[29] Wu, K., Zhang, Q. & Poon, A. Integrated Si 3 N 4 microresonator-based quantum light sources with high brightness using a subtractive wafer-scale platform. *Optics Express.* **29**, 24750-24764 (2021)

[30] Ji, X., Barbosa, F., Roberts, S., Dutt, A., Cardenas, J., Okawachi, Y., Bryant, A., Gaeta, A. & Lipson, M. Ultra-low-loss on-chip resonators with sub-milliwatt parametric oscillation threshold. *Optica.* **4**, 619-624 (2017)

[31] Xiang, C., Liu, J., Guo, J., Chang, L., Wang, R., Weng, W., Peters, J., Xie, W., Zhang, Z., Riemensberger, J. & Others Laser soliton microcombs heterogeneously integrated on silicon. *Science.* **373**, 99-103 (2021)

[32] Xuan, Y., Liu, Y., Varghese, L., Metcalf, A., Xue, X., Wang, P., Han, K., Jaramillo-Villegas, J., Al Noman, A., Wang, C. & Others High-Q silicon nitride microresonators exhibiting low-power frequency comb initiation. *Optica.* **3**, 1171-1180 (2016)

[33] Jin, W., Yang, Q., Chang, L., Shen, B., Wang, H., Leal, M., Wu, L., Gao, M., Feshali, A., Paniccia, M. & Others Hertz-linewidth semiconductor lasers using CMOS-ready ultra-high-Q microresonators. *Nature Photonics.* **15**, 346-353 (2021)

[34] Dutt, A., Luke, K., Manipatruni, S., Gaeta, A., Nussenzveig, P. & Lipson, M. On-chip optical squeezing. *Physical Review Applied.* **3**, 044005 (2015)

[35] Stern, B., Ji, X., Okawachi, Y., Gaeta, A. & Lipson, M. Battery-operated integrated frequency comb generator. *Nature.* **562**, 401-405 (2018)

[36] Shen, B., Chang, L., Liu, J., Wang, H., Yang, Q., Xiang, C., Wang, R., He, J., Liu, T., Xie, W. & Others Integrated turnkey soliton microcombs. *Nature.* **582**, 365-369 (2020)

[37] Raja, A., Voloshin, A., Guo, H., Agafonova, S., Liu, J., Gorodnitskiy, A., Karpov, M., Pavlov, N., Lucas, E., Galiev, R. & Others Electrically pumped photonic integrated soliton microcomb. *Nature Communications.* **10**, 1-8 (2019)

[38] Thew, R., Nemoto, K., White, A. & Munro, W. Qudit quantum-state tomography. *Physical Review A.* **66**, 012303 (2002)

[39] Zhong, H., Wang, H., Deng, Y., Chen, M., Peng, L., Luo, Y., Qin, J., Wu, D., Ding, X., Hu, Y. & Others Quantum computational advantage using photons. *Science.* **370**, 1460-1463 (2020)

**Acknowledgements.** — H. M., R. J., A. K. K., R. H., and M. K. acknowledge funding from the German federal ministry of education and research, Quantum Futur Program (PQuMAL), from European Research Council (ERC) under the European Union's Horizon 2020 research and innovation programme under grant agreement No. 947603 (QFreC project), and from Deutsche Forschungsgemeinschaft (DFG, German Research Foundation) under Germany's Excellence Strategy within the Cluster of Excellence PhoenixD (EXC 2122, Project ID 390833453). R. H. acknowledges the financial support provided by the Alexander von Humboldt Stiftung to conduct the research. A.V.R., J. E., and K. -J. B. acknowledge funding from the EU within the project 3PEAT.

**Author contributions.** — H. M. developed the initial setup for correlation measurements. A. K. K., R. J., and R. H. constructed and tested the RF-modulation setup to perform the tomography and Bell test. A. V. R., J. P. E., and K.-J. B. supervised the device-design, and the chip was fabricated, and assembled by Lionix Intl. H. M. characterized the device and analyzed the data with R. J. and R. H.. M. K. supervised the overall project. All authors discussed the results and contributed to the writing of the manuscript.

# Methods: Fully on-chip photonic turnkey quantum source for entangled qubit/qudit state generation

Hatam Mahmudlu,[1,2,3] Robert Johanning,[1,2,3] Anahita Khodadad Kashi,[1,2,3] Albert van Rees,[4] Jörn P. Epping,[5,6] Raktim Haldar,[1,2,3,*] Klaus-J. Boller,[4] and Michael Kues[1,2,3,†]

[1]*Institute of Photonics, Leibniz University Hannover, Nienburger Straße 17, 30167 Hannover, Germany*
[2]*Hannover Centre for Optical Technologies, Leibniz University Hannover, Nienburger Straße 17, 30167 Hannover, Germany*
[3]*Cluster of Excellence PhoenixD (Photonic, Optics, and Engineering – Innovation Across Disciplines), Leibniz University Hannover, Hannover, Germany*
[4]*Laser Physics and Nonlinear Optics, Mesa+Institute for Nanotechnology, Department for Science and Technology, University of Twente, PO Box 217, Enschede, 7500 AL, The Netherlands*
[5]*QuiX Quantum, 7521, AN Enschede, The Netherlands*
[6]*Lionix Intl., 7521, AN Enschede, The Netherlands*

In this document, at first, we describe the methods regarding the fabrication process of this InP-Si$_3$N$_4$ hybrid turnkey quantum photonic source. Then we provide a comprehensive details on the coincidence measurement techniques, theoretical analysis and experimental procedures of quantum state tomography, and Bell test measurements with other relevant information.

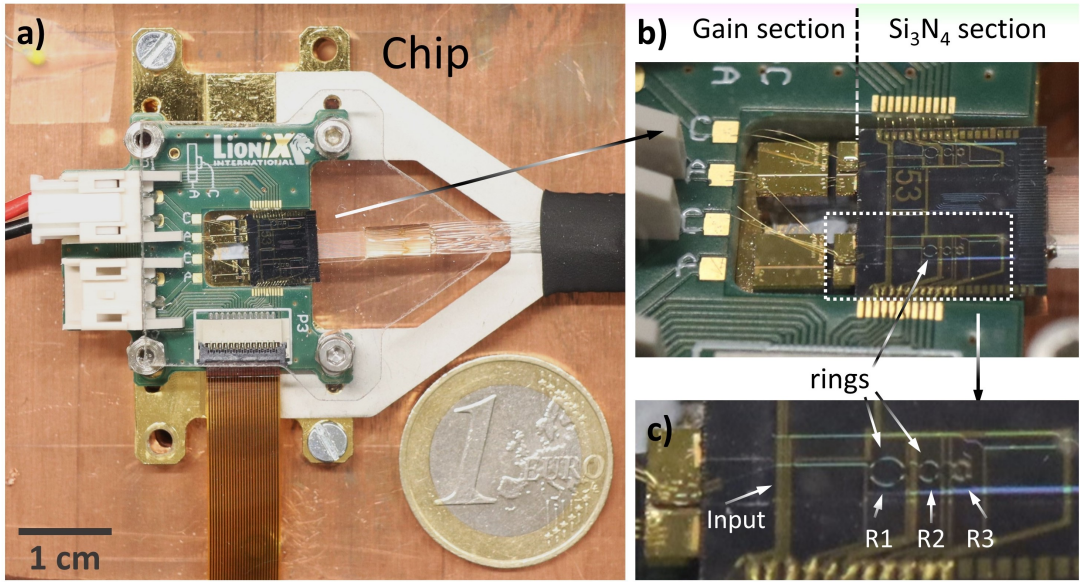

FIG. E.1. **Photographs of the designed chip. a)** Size of the laser-integrated chip is compared to a one Euro coin. **b)** Enlarged photograph, both the InP gain section and the rest of the circuit, i.e., where the Si$_3$N$_4$ Vernier filtering and the SFWM photon-pair generation process occured, are marked. **c)** Further magnified image, where the three rings R1, R2, and R3, are visible. The perimeters of R1 and R2 are 1579 µm, and 1005 µm, respectively, whereas the entangled photon pair generation took place in the smallest ring R3 having a circumference of about 850 µm.

**Design and fabrication of the hybrid photonic quantum source.** The design of the laser-integrated hybrid photonic source is schematically shown in Fig. 1a in the main text. It comprises a III-V reflective semiconductor optical amplifier (RSOA), which is hybrid integrated with a Si$_3$N$_4$-based feedback circuit. To provide optical gain around the central wavelength of 1550 nm, a standard RSOA fabricated by Fraunhofer HHI is used, that contains a 700 µm long InP-based quantum well amplifier. One mirror of the laser cavity is formed by a high reflective coating ($R = 0.90$) on the end facet of this amplifier. For the realization of the feedback circuit, we chose Si$_3$N$_4$-based TriPleX waveguides with an asymmetric double stripe cross-section as described in[1]. This waveguide cross-section supports single TE$_{00}$ transversal mode propagation with low propagation losses ($< 0.1$ dB/cm) and low bend loss at tight bend radii (down to 100 µm), which enables the design of a sharp spectral filter with high Finesse and large free spectral range. The spectral filter on this feedback circuit is formed by three consecutive microring resonators[2–5]. These rings have circumferences of 1579 µm, 1005 µm, and 850 µm and corresponding free spectral ranges (FSRs) of 107 GHz, 169 GHz, and 199 GHz, respectively. The designed bus-to-ring power coupling coefficient is 0.03. The Vernier filter selects a single frequency and suppresses the ASE background from the RSOA over a bandwidth of 15.3 THz, which covers the entire gain bandwidth. The second mirror of the laser cavity is formed by a Sagnac loop after a balanced and tunable Mach Zehnder interferometer (MZI), to reflect and extract an adjustable fraction of the intra-cavity light. Light can be coupled out from the cavity at any of the ports depicted in Fig. 1a. All ports are coupled to an array of single-mode polarization-maintaining fibers. To maximize the coupling efficiency, and to prevent undesired feedback from the interfaces between the gain chip, feedback chip and fibers, several measures have been taken including waveguide tapering and tilting[4]. For thermal tuning of the MRRs, the MZI and an extra phase shifter for the cavity length, resistive heaters are placed on top of these elements. The fabrication of the feedback chips was carried out with a process flow as described in[1]. After wafer processing, gain chip, feedback chip and fibers were aligned and permanently fixed together with adhesive bonding. The heaters and amplifier were wire bonded to a printed circuit board for convenient electronic control. Finally, the optical chips and fibers, and the corresponding electronics were mechanically fixed on a common mount to form a single and stable assembly for the hybrid photonic quantum source.

**Miniaturization factor.** A photograph of the electrically-pumped laser-integrated qubit/qudit source is shown in Fig. E.1a. The entire chip size is less than one Euro coin. The InP gain-section and rest of the circuit consisting of Si$_3$N$_4$ Vernier filtering and the SFWM photon-pair generation stages are shown in the enlarged Fig. E.1b with the three rings, viz. R1, R2, and R3, visible prominently in Fig. E.1c. This is so far the first known scalable quantum photonic source with laser and Vernier filter integrated on-chip, which generates with a very high generation rate near-perfect frequency-bin entangled high-dimensional Bell-states. Compared to similar studies[6], which used a nonlinear chip, external laser, and a filter occupying roughly a space of two to three shoeboxes or about one-third of a standard optical bench, this chip is at least $(300 \times 200)$ times smaller. Further, due to the light-weight of this chip, it is perfect for out-of-the-lab applications, e.g., quantum satellite communications.

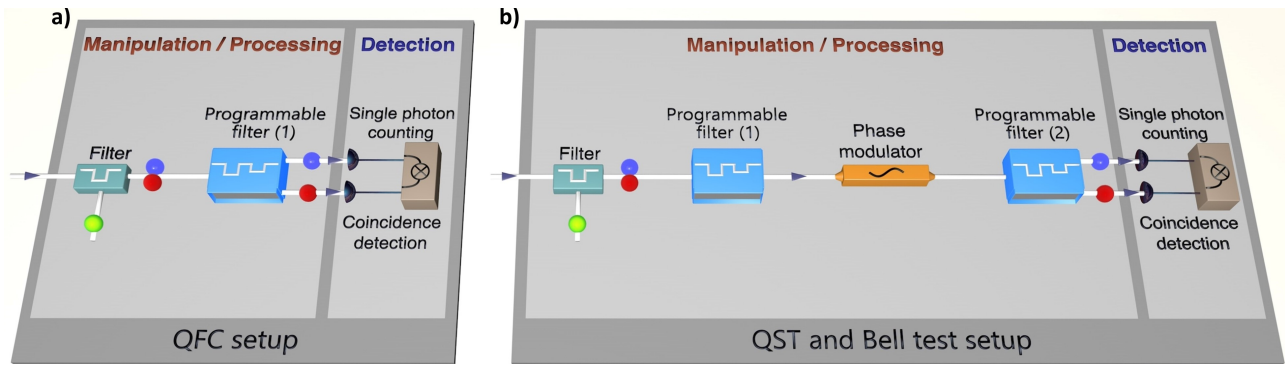

FIG. E.2. **a) Experimental setup for the quantum frequency comb (QFC) characterization.** The signal and idler photons, being superposition of the frequency modes defined by the resonances of QFCs, are collected from the chip. After passing the high rejection notch filter (NF) and programmable filter (PF) the photon pairs were detected by two superconducting nanowire single-photon detectors (SNSPDs) prior to determining their time correlations by a time-tagger (TT). **b) Experimental setup for the frequency-bin entanglement characterization.** Programmable filters (1) and (2) and an electro-optic phase modulator (EOM) were used to manipulate and characterize the frequency entangled quantum states before the signal and idler photons were detected by two SNSPDs and the correlation measured by the time-tagger.

**Experimental setup for the quantum frequency comb (QFC) emission characterization.** A schematic of the QFC characterization experiments determining the frequency-time correlation of the photon-pairs is shown in Fig. E.2a. The idler and the signal photons generated through a SFWM process in the third MRR, exhibit correlations in the frequency-time domain over several frequency modes due to energy $(2\omega_{\rm p} = \omega_{\rm s} + \omega_{\rm i})$ and momentum conservation $(2\kappa_{\rm p} = \kappa_{\rm s} + \kappa_{\rm i})$[7]. These photon pairs (see Fig. E.1a) were out-coupled from the chip and transmitted to a high rejection notch filter (NF) with an isolation bandwidth of 200 GHz and an out-of-band suppression of 100 dB in order to separate the excitation field from the single/idler photons. Next, the signal and idler photons were split into different spatial modes and the photons frequency components were selected by a programmable filter with a bandwidth of 5 GHz for each photon. Subsequently, the photon pairs at the selected comb modes were detected by two superconducting nanowire single photon detectors (SNSPDs). Finally, the time correlations between the photons at the output of the two channels were determined by a time tagger (TT), working in a 5 ps resolution mode.

**Experimental setup for frequency-bin entanglement characterization.** For the quantum state tomography and the Bell test, it is necessary to manipulate the photonic states by mixing the individual frequency modes to realize quantum interference and perform projection measurements. The setup two for this is shown in Fig. E.2b. The photon pairs were sent to a first (leftmost) programmable filter (1), where suitable attenuation- and phase masks were applied to the QFC, specifically selecting with a 25 GHz bandwidth the corresponding frequency modes. As shown in Fig. 2a of the main manuscript, for Bell-tests and QST on qubits we have chosen frequency modes 2 and 3, as well as 3 and 4. For the same measurements on qutrits we have selected modes 2, 3 and 4.

Then the photons were passed through an electro-optic phase modulator (EOM) driven by a single sinusoidal radio-frequency (RF) tone. Sidebands created through this process allowed the mixing of the frequency modes. It is a crucial prerequisite to perfectly adjust the EOM driving signal in such a way that integer values of its driving frequency match the FSR of ring R3 (i.e., 198.9 GHz). The RF-source amplifier combination used to drive the EOM in our experiments had a maximum RF bandwidth of 25 GHz, which allowed us to adjust it to 24.863 GHz providing an integer ring FSR to EOM frequency ratio of eight. Specifically, the $8^{\rm th}$ sideband is required to mix a resonance with its neighboring mode(s). The EOM frequency adjustment and the phase offset compensation (EOM offset compensation) procedures are described in the next paragraphs.

The amplitude of the $n^{\rm th}$ sideband at a particular RF power can be described by the $n^{\rm th}$ order Bessel function of the first kind, where the RF power is the argument of the Bessel function. For measurements on the two-dimensional (2D) state, we set the RF signal power to $-7.5$ dBm, where the amplitude of the $4^{\rm th}$ sideband was maximal. This setting caused a mixing of two adjacent modes $|\bar{k}\rangle$, and $|\bar{k}+1\rangle$ in a frequency mode centered between the two frequency modes. For projection measurements on the three-dimensional state, we applied a signal power of $-4.85$ dBm to the EOM, at which the $0^{\text{th}}$ and the $8^{\text{th}}$ sideband were created with the same amplitude. This was set by comparing the single count rates of these sidebands for different values of the EOM power. As a result, the modes $|\bar{k}\rangle$, and $|\bar{k}+2\rangle$ were mixed to the frequency mode $|\bar{k}+1\rangle$, which itself remained at this position with the same amplitude as the other two. As the intensity of the sidebands created by the EOM decreased with the order of the band we were currently limited to perform entanglement measurements and successful tomography for a dimension of three. In future, this limitation can be circumvented by using another RF-drive with a different, e.g., 50 GHz driving frequency.

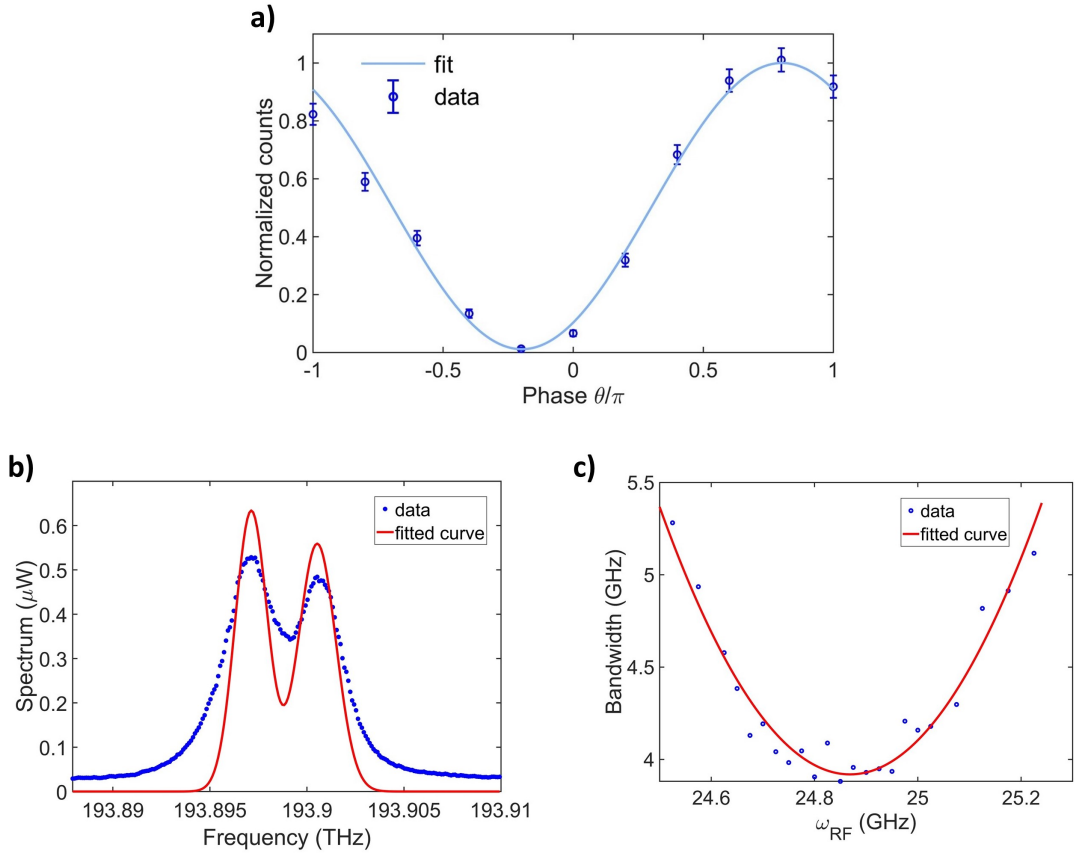

FIG. E.3. **Adjustment of the EOM. a)** Interference measurement of the EOM induced offset. The EOM mixed two frequency modes and the phase of one mode was continuously changed from 0 to $2\pi$ via PF1 while coincidence counts were measured. The maximum of the curve corresponds to a superposition without a resulting phase difference between mixed modes, so that this phase was used as a reference. **b)** Spectrum of two modes mixed by the EOM with a mismatch in FSR and $\omega_{\text{RF}}$. The FWHM of this curve was determined. **c)** Interpolation of the FWHM of the mixed modes for different $\omega_{\text{RF}}$ to find the optimum.

At last, a second programmable filter (2) collected the photons within a bandwidth of 25 GHz and routed the individual signal/idler photons to the respective SNSPDs for coincidence measurements with the time tagger. The coincidence measurement time was 5 min per setting for the quantum interference of two-dimensional states and 10 min per setting for the interference of three-dimensional states as well as for all tomography measurements.

Due to the resonance linewidth of the microring R3, the dispersion of the waveguides, and the timing jitter of the detectors, the time difference between the detection of the two photons of a coincidence event was distributed in a histogram following a Lorentz curve. We selected those counts that lie in a predefined symmetric window around the center.

**Setting of EOM driving frequency.** For optimal quantum interference and projection measurements, perfect overlap of the mixed modes is required. This was achieved by setting the EOM driving frequency to a multiple of the FSR. We used an erbium-doped fiber amplifier (EDFA) as a broadband light source to obtain the transmission spectra of the MRR R3. Next, the transmitted signal of the ring (see Fig. 1a) passed the EOM and was then sent to a wave analyzer, where the FWHM of the mixed sidebands was measured for different EOM driving frequencies (Fig. E.3b). As a result, the interpolated minimum of the measured FWHM values defined the RF frequency providing the maximum possible overlap for the modes (Fig. E.3c).

**EOM offset compensation.** Driving the EOM introduces an additional phase to the sidebands, which depends on the RF-source's power and frequency and scales proportional to the order of the sidebands. We measured this phase between two resonant modes by creating the superposition of both signal- and both idlers and rotating the phase of only one mode in the first programmable filter (PF1) (see Fig. E.2b). Thereby, the measured coincidence counts follow a sine curve, where its maximum marks the phase of PF1 that compensates for the phase of the EOM. Once this phase was determined, it was kept fixed for the entire measurement. This was done for the combination of modes 2 and 3 as well as modes 3 and 4 for the two-dimensional states, resulting in phases of $+0.84\,\mathrm{rad}$ and $-0.51\,\mathrm{rad}$ for mode 3 and mode 4, respectively (Fig. E.3a). For three-dimensional state characterization, the phases of the resonances 2 and 4 were both corrected with respect to the resonant mode 3 at PF1.

**Quantum interference measurement.** We performed a dedicated Bell test measurement for both the two- and three-dimensional states. This kind of measurement can give an indication of the violation of classical Bell inequality and is used extensively to prove the presence of entanglement[8].

We used resonance modes 2, 3, and 4, which had approximately the same count rates. By mixing two or three of these resonant modes, we thus expected to measure a Bell state of the form:

$$|\psi\rangle = \sum_{k=2}^{4} c_k |k\rangle_s |k\rangle_i \approx \frac{1}{\sqrt{3}} \sum_{k=2}^{4} |k\rangle_s |k\rangle_i \tag{E.1}$$

Starting from this, we created the following projections for signal and idler by mixing all present signal- and idler modes, respectively, and applying phases as follows:

$$|\psi_{\mathrm{proj},2}\rangle = \frac{1}{\sqrt{2}} \left( |\bar{k}\rangle + e^{i\theta} |\bar{k}+1\rangle \right) \tag{E.2}$$

$$|\psi_{\mathrm{proj},3}\rangle = \frac{1}{\sqrt{3}} \left( e^{i\theta} |\bar{k}\rangle + |\bar{k}+1\rangle + e^{i2\theta} |\bar{k}+2\rangle \right) \tag{E.3}$$

With $\bar{k} = 2$ (2D and 3D) and $\bar{k} = 3$ (only 2D). We rotated the phase $\theta$ from 0 to $\pi$ for signal and idler in steps of $\pi/10\,\mathrm{rad}$ and measured the coincidence counts. The analytical expression for the intensities of the interfered signal as function of the phase $\theta$ can be given by

$$C_2\left(\theta\right) = 1 + \epsilon_2 \cos(2\theta) \tag{E.4}$$

$$C_3\left(\theta\right) = 3 + 2\epsilon_3 \left[2\cos(2\theta) + \cos(4\theta)\right] \tag{E.5}$$

For the fitting of these curves, we used the maximum of the noise background from the set of measurements as a lower bound of the curve. We were able to reach this limit because of the meticulous synchronization of the EOM frequency to the FSR, enabling perfect destructive interference. Further, we did not subtract the background. From the parameter $\epsilon_D$, which measures the deviation from the maximally entangled state, the visibility $V_D$ can be derived as,

$$V^{(2)} = \epsilon_2 \tag{E.6}$$

$$V^{(3)} = \frac{3\epsilon_3}{2 + \epsilon_3} \tag{E.7}$$

Violation of the Bell inequality is achieved for $\frac{1}{\sqrt{2}} \approx 0.7071 < V^{(2)}$ and $\frac{3(6\sqrt{3}-9)}{6\sqrt{3}-5} \approx 0.7746 < V^{(3)}$ which is a proof for the presence of entanglement[8]. We obtained values of $V_{2-3}^{(2)} = 0.975 \pm 0.002$ and $V_{3-4}^{(2)} = 0.976 \pm 0.002$ for qubits and $V_{2-3-4}^{(3)} = 0.968 \pm 0.004$ for qutrits, violating the Bell inequality. Note that here the superscript of the symbol within first bracket denotes the dimension of the quantum state and the subscript represents the numbers of the frequency-bins on which the tomography was performed.

**Density matrix reconstruction.** We use quantum state tomography to reconstruct the density matrix from a tomographically complete set of projection measurements (quorum), described by projection wavevectors $|\psi_\nu\rangle$[9]. Thereby we expect the coincidence counts to be

$$n_v = C \langle \psi_v | \hat{\rho} | \psi_v \rangle \tag{E.8}$$

for a constant $C$ depending on the measurement time, pair generation rate, losses etc. Using our experimental data as $n_\nu$, we reconstruct $\hat{\rho}$ using the relations[8,9],

$$\hat{\rho} = C^{-1} \sum_v M_v n_v \tag{E.9}$$

$$M_v = \sum_x \Gamma_x \left(B^{-1}\right)_{x,v} \tag{E.10}$$

$$B_{x,y} = \langle \psi_x | \Gamma_y | \psi_x \rangle \tag{E.11}$$

$$C = \sum_k n_k \text{ for } tr\{M_k\} = 1 \tag{E.12}$$

For 2D we applied the following projections: $|\bar{k}\rangle$, $|\bar{k}+1\rangle$, $\frac{1}{\sqrt{2}}\left(|\bar{k}\rangle + |\bar{k}+1\rangle\right)$, $\frac{1}{\sqrt{2}}\left(|\bar{k}\rangle + i\,|\bar{k}+1\rangle\right)$ for signal and idler, respectively, resulting in a total of 16 projection measurements.

For the 3D QST[10], the method given in the last paragraph was extended and we applied the projections

$$\frac{1}{\sqrt{2}}\left(|\bar{k}\rangle + |\bar{k}+1\rangle\right) \tag{E.13}$$

$$\frac{1}{\sqrt{2}}\left(e^{\frac{2\pi}{3}i}\,|\bar{k}\rangle + e^{-\frac{2\pi}{3}i}\,|\bar{k}+1\rangle\right) \tag{E.14}$$

$$\frac{1}{\sqrt{2}}\left(e^{-\frac{2\pi}{3}i}\,|\bar{k}\rangle + e^{\frac{2\pi}{3}i}\,|\bar{k}+1\rangle\right) \tag{E.15}$$

$$\frac{1}{\sqrt{2}}\left(|\bar{k}\rangle + |\bar{k}+2\rangle\right) \tag{E.16}$$

$$\frac{1}{\sqrt{2}}\left(e^{\frac{2\pi}{3}i}\,|\bar{k}\rangle + e^{-\frac{2\pi}{3}i}\,|\bar{k}+2\rangle\right) \tag{E.17}$$

$$\frac{1}{\sqrt{2}}\left(e^{-\frac{2\pi}{3}i}\,|\bar{k}\rangle + e^{\frac{2\pi}{3}i}\,|\bar{k}+2\rangle\right) \tag{E.18}$$

$$\frac{1}{\sqrt{2}}\left(|\bar{k}+1\rangle + |\bar{k}+2\rangle\right) \tag{E.19}$$

$$\frac{1}{\sqrt{2}}\left(e^{\frac{2\pi}{3}i}\,|\bar{k}+1\rangle + e^{-\frac{2\pi}{3}i}\,|\bar{k}+2\rangle\right) \tag{E.20}$$

$$\frac{1}{\sqrt{2}}\left(e^{-\frac{2\pi}{3}i}\,|\bar{k}+1\rangle + e^{\frac{2\pi}{3}i}\,|\bar{k}+2\rangle\right) \tag{E.21}$$

Again, the surplus modes for signal and idler, which were not required for the specific projection, were blocked by the PF1. No count adjustment/correction was needed, as every projection we used consists of exactly two modes for signal and idler, respectively.

In reality, a density matrix corresponding to a physical quantum state has to be Hermitian and positive semidefinite. Due to experimental imperfections, this is usually not the case for a reconstructed matrix. This is treated by applying a maximum likelihood estimation, which is a numerical optimization finding the closest physical density matrix relative to the measurement data[9].

**Fidelity.** Fidelity is a measure of the distance between two quantum states, where a unit fidelity implicates the two states are perfectly identical. For two density matrices $\hat{\rho}_1$ and $\hat{\rho}_2$, the fidelity reads as follows:

$$F\left(\hat{\rho}_1, \hat{\rho}_2\right) = \left(\mathrm{tr}\sqrt{\sqrt{\hat{\rho}_1}\hat{\rho}_2\sqrt{\hat{\rho}_1}}\right)^2 \tag{E.22}$$

We calculated the fidelities from the measured quantum states with respect to the pure 2D and 3D Bell states as $F_{2-3}^{(2)} = (0.988 \pm 0.01)$ and $F_{3-4}^{(2)} = (0.981 \pm 0.01)$ for qubit and $F_{2-3-4}^{(3)} = (0.806 \pm 0.02)$ for qutrit, respectively. The differences are explained by the lower coincidence rates for 3D than for 2D, as the intensity of the created EOM sidebands decreases in general with the order ($4^\mathrm{th}$ vs. $8^\mathrm{th}$), in turn increasing the noise from up to $0.1\,\%$ for qubits to $6\,\%$ for qutrits, as predicted by the linear model of uncolored noise[11]. Taking this noise into account yields a fidelity of $F_{2-3-4}^{(3),\,\mathrm{noise}} = (0.857 \pm 0.02)$, while the improvement for the qubit case was only up to 0.001. In addition, the number of measurements required for 3D QST is 81, significantly higher than the 16 measurements for the 2D case, thereby making 3D measurements more prone to be exposed to the laser instabilities due to the much longer measurement time.

**Bell inequality.** The Bell test measurements are directly related to the $S$-parameters of Bell inequalities by the multiplication of the parameter $\epsilon_\mathrm{D}$ with the respective maximal value of the inequality, which is $2\sqrt{2} \approx 2.8284$ for qubits and $4/\left(6\sqrt{3} - 9\right) \approx 2.8729$ for qutrits[11]. Our measurements yield values of $S = 2.758$, and $S = 2.760$ for 2D and $S = 2.735$ for 3D, where $S > 2$ means a violation of the respective inequality. We can also derive the $S$-parameters from the reconstructed density matrices, yielding values of up to $S = 2.793$, which is in good agreement with the Bell tests. Just as with the Bell tests, the violation of a Bell inequality proves the presence of entanglement.

---

\* raktim.haldar@iop.uni-hannover.de

† michael.kues@iop.uni-hannover.de

1 Roeloffzen, C., Hoekman, M., Klein, E., Wevers, L., Timens, R., Marchenko, D., Geskus, D., Dekker, R., Alippi, A., Grootjans, R. & Others Low-loss Si3N4 TriPleX optical waveguides: Technology and applications overview. *IEEE Journal Of Selected Topics In Quantum Electronics.* **24**, 1-21 (2018)

2 Matsumoto, T., Suzuki, A., Takahashi, M., Watanabe, S., Ishii, S., Suzuki, K., Kaneko, T., Yamazaki, H. & Sakuma, N. Narrow spectral linewidth full band tunable laser based on waveguide ring resonators with low power consumption. *Optical Fiber Communication Conference.* pp. OTh5Q5 (2010)

3 Tran, M., Huang, D., Guo, J., Komljenovic, T., Morton, P. & Bowers, J. Ring-resonator based widely-tunable narrow-linewidth Si/InP integrated lasers. *IEEE Journal Of Selected Topics In Quantum Electronics.* **26**, 1-14 (2019)

4 Fan, Y., Rees, A., Slot, P., Mak, J., Oldenbeuving, R., Hoekman, M., Geskus, D., Roeloffzen, C. & Boller, K. Hybrid integrated InP-Si3N4 diode laser with a 40-Hz intrinsic linewidth. *Optics Express.* **28**, 21713-21728 (2020)

5 Stern, B., Ji, X., Okawachi, Y., Gaeta, A. & Lipson, M. Battery-operated integrated frequency comb generator. *Nature.* **562**, 401-405 (2018)

6 Xuan, Y., Liu, Y., Varghese, L., Metcalf, A., Xue, X., Wang, P., Han, K., Jaramillo-Villegas, J., Al Noman, A., Wang, C. & Others High-Q silicon nitride microresonators exhibiting low-power frequency comb initiation. *Optica.* **3**, 1171-1180 (2016)

7 Li, X., Voss, P., Sharping, J. & Kumar, P. Optical-fiber source of polarization-entangled photons in the 1550 nm telecom band. *Physical Review Letters.* **94**, 053601 (2005)

8 Kues, M., Reimer, C., Roztocki, P., Cortés, L., Sciara, S., Wetzel, B., Zhang, Y., Cino, A., Chu, S., Little, B. & Others On-chip generation of high-dimensional entangled quantum states and their coherent control. *Nature.* **546**, 622-626 (2017)

9 James, D., Kwiat, P., Munro, W. & White, A. Measurement of qubits. *Phys. Rev. A.* **64**, 052312 (2001,10)

10 Thew, R., Nemoto, K., White, A. & Munro, W. Qudit quantum-state tomography. *Physical Review A.* **66**, 012303 (2002)

11 Collins, D., Gisin, N., Linden, N., Massar, S. & Popescu, S. Bell inequalities for arbitrarily high-dimensional systems. *Physical Review Letters.* **88**, 040404 (2002)